Title: On the Melnikov method for fractional-order systems

By Authors: Hang Li, Yongjun Shen, Jian Li, Jinlu Dong, Guangyang Hong

# On the Melnikov method for fractional-order systems

Hang Li [1], Yongjun Shen [2, 3], Jian Li [1,*], Jinlu Dong [1], Guangyang Hong [1,*]

[1] Key Laboratory of Structural Dynamics of Liaoning Province, College of Sciences, Northeastern University, Shenyang, 110819, China.

[2] Department of Mechanical Engineering, Shijiazhuang Tiedao University, Shijiazhuang, 050043, China.

[3] State Key Laboratory of Mechanical Behavior and System Safety of Traffic Engineering Structures, Shijiazhuang Tiedao University, Shijiazhuang, 050043, China.

[*] Corresponding author. Email: jianli@mail.neu.edu.cn (J. Li), hongguangyang@mail.neu.edu.cn (G. Hong).

**Abstract**

This paper is dedicated to clarifying and introducing the correct application of Melnikov method in fractional dynamics. Attention to the complex dynamics of hyperbolic orbits and to fractional calculus can be, respectively, traced back to Poincaré's attack on the three-body problem a century ago and to the early days of calculus three centuries ago. Nowadays, fractional calculus has been widely applied in modeling dynamic problems across various fields due to its advantages in describing problems with non-locality. Some of these models have also been confirmed to exhibit hyperbolic orbit dynamics, and recently, they have been extensively studied based on Melnikov method, an analytical approach for homoclinic and heteroclinic orbit dynamics. Despite its decade-long application in fractional dynamics, there is a universal problem in these applications that remains to be clarified, i.e., defining fractional-order systems within finite memory boundaries leads to the neglect of perturbation calculation for parts of the stable and unstable manifolds in Melnikov analysis. After clarifying and redefining the problem, a rigorous analytical case is provided for reference. Unlike existing results, the Melnikov criterion here is derived in a globally closed form, which was previously considered unobtainable due to difficulties in the analysis of fractional-order perturbations characterized by convolution integrals with power-law type singular kernels. Finally, numerical methods are employed to verify the derived Melnikov criterion. Overall, the clarification for the problem and the presented case are expected to provide insights for future research in this topic.

**Keywords:** Melnikov method; Fractional dynamics; Fractional-order systems; Horseshoe Chaos; Chaos threshold; Homoclinic and Heteroclinic orbit.

## 1 Introduction

In 1890, Poincaré's global qualitative analysis on the three-body problem [1] marked a turning point in dynamics. With this work, the complexity of hyperbolic orbits was noticed for the first time. As he later remarked [2], "*One will be struck by the complexity of this figure, which I do not even attempt to draw. Nothing better illustrates the complexity of the three-body problem and, in general, all dynamic problems where there is no single-valued integral and Bohlin's series diverge.*" Subsequent significant progress on this issue occurred in 1963, Melnikov [3] considered a general planar dynamical system perturbed by a small periodic perturbation, when he ingeniously developed a global perturbation technique. The technique is able to determine the existence of a transverse intersection between the stable and unstable manifolds of the perturbed splitting hyperbolic homoclinic orbit, by using the globally computable solution of the unperturbed system. According to the Smale-Birkhoff homoclinic theorem [4], such transverse intersections are the signature of the existence of the horseshoe map and its associated chaotic dynamics. This global perturbation technique seemed to be little known for a period of time until it was rediscovered by Holmes [5] in 1979 and applied to the analytical prediction of chaotic dynamics of the Duffing oscillator. Nowadays, this global perturbation technique is well known as Melnikov method, and it has been widely applied to the analytical prediction of chaotic solution of nonlinear systems [6-9].

Another timeline can be traced back to an earlier period, in 1695, the early days of classical calculus theory, when Leibniz and l'Hôpital discussed in a correspondence the possibility of extending Leibniz's derivative notation $\mathrm{d}^n y / \mathrm{d}x^n$ to the order 1/2. Leibniz replied, "*It will lead to a paradox, from which one day useful consequences will be drawn.*" This correspondence is accepted as the origin of fractional calculus, and the problem mentioned has, in fact, continued to attract the interest of mathematicians for a long time thereafter. At the end of one of his works [10] in 1729, Euler wrote, "*I want to add something that is more interesting* *than useful…; For positive integers n,* $\mathrm{d}^n$ *can be found by repeated differentiation. However, for fractional n, it is not applicable. Nevertheless, by considering the interpolation of the series discussed in this paper, it is possible to address this problem…*" Thereafter, Lagrange in 1772, Laplace in 1812, Fourier in 1822, Riemann in 1847, Grünwald in 1867, Letnikov in 1868, Weyl in 1919, and others (see Ref. [11] for a summary), have directly or indirectly contributed to the advance of fractional calculus theory. For over 300 years, research on fractional calculus was primarily in pure mathematics. It was not until the last 30 years, as Leibniz predicted, that fractional calculus began to be applied to various fields such as control engineering [12-13], physics [14], mechanical systems and signal processing [15-16]. During this period, numerical methods of fractional calculus [17-19] were gradually developed, promoting the application in other fields. For more applications of fractional calculus, one can refer to the recent reviews by Sun et al. [20] and Diethelm et al. [21].

In particular, in some dynamic problems modeled by fractional calculus, the dynamics of hyperbolic orbits have also been confirmed. For the purpose of utilizing or suppressing chaos, a naturally arising idea is to continue using the Melnikov method to analyze the global dynamics of these fractional-order systems to predict their chaotic thresholds. This idea has been widely practiced since 2015 in various dynamic problems [22-35], especially in mechanical vibration problems [22-25], providing useful theoretical insights for the problems they describe. However, despite being applied for a decade, there is a universal omission in these applications regarding the analysis of fractional-order perturbations that needs to be clarified. Specifically, the definition of fractional-order governing equations on finite memory boundaries leads to incomplete perturbation calculations in Melnikov analysis. One of the main works of this paper is to clarify this point and redefine the problem.

On the other hand, a brief review of existing Melnikov analysis work on fractional-order systems in this paper reveals that The Melnikov criteria for these systems are usually not obtained in a global closed form but rather in a semi-analytical [22-27] or local analytical [28-35] manner. After summarizing and categorizing these results and redefining the problem, the fractional-order Duffing-Rayleigh system is used as an example to re-demonstrate the Melnikov analysis for fractional-order systems, obtaining its Melnikov criteria in a global closed form.

The presented analysis and results are expected to provide a reference for future research on Melnikov analysis in fractional dynamics.

The remainder of this paper is organized as follows. In Section 2, the application of the Melnikov method in fractional dynamics over the past decade is briefly reviewed, summarized, and categorized. Then, by combining the basic idea of Melnikov method and the memory principle of time fractional-order derivatives, it is clarified how to define the problem and then perform a complete perturbation analysis. In Section 3, the fractional-order Duffing-Rayleigh system is used as an example to re-demonstrate the Melnikov analysis for reference. In Section 4, the main work and conclusions of this paper are summarized.

## 2 Summary and clarification on the problem

In this paper, we concentrate on the Melnikov method for the following time-periodic system in the case with fractional-order elements.

$$\dot{\boldsymbol{x}} = \boldsymbol{f}(\boldsymbol{x}) + \varepsilon \boldsymbol{g}(\boldsymbol{x}, t) \tag{1}$$

where $\boldsymbol{x} = [x_1, x_2]^{\mathrm{T}} \in \mathbb{R}^2$, $\boldsymbol{f} = [f_1, f_2]^{\mathrm{T}} : \mathbb{R}^2 \to \mathbb{R}^2$ and similarly for $\boldsymbol{g}$. The vector field $\boldsymbol{f}$ is Hamiltonian, whereas $\boldsymbol{g}$ need not be. To facilitate the subsequent discussion of the technical details, the conventional Melnikov method is thoroughly introduced in Appendix A, and the definitions of fractional calculus are introduced in Appendix B.

As aforementioned, there is a universal omission in the existing Melnikov analysis of fractional-order systems that requires clarification. Additionally, in these works, the Melnikov criteria are generally not derived in a globally closed form. Motivated by these two issues, the following work is dedicated to reviewing and summarizing the existing works [22-35] first, then clarifying how to correctly perform global perturbation calculations on fractional-order systems, and finally demonstrating a rigorous analytical case to provide insights for future research on this topic.

### 2.1 A minor review on 10 years of research

Specifically, some of the modeled fractional-order systems contain hyperbolic orbits, so that it possible to analyze the global dynamics based on Melnikov method. Here, it is assumed that they still are or can be rewritten in the form of Eq. (1) and contain at least one fractional-order element to constitute the following perturbation.

$$\boldsymbol{g}\left(\boldsymbol{x},t\right)=\boldsymbol{g}_{I}\left(\boldsymbol{x},t\right)+\boldsymbol{g}_{F}\left(\mathrm{D}^{q}\boldsymbol{x},\boldsymbol{x},t\right) \tag{2}$$

Note that in the existing literature, only the case $0<q<2$, which is common in vibration problems of mechanical systems, is studied, and will be so in this paper. In this case, the reason why the fractional-order element is considered here as perturbation term placed in $\boldsymbol{g}\left(\boldsymbol{x},t\right)$ lies in its non-conservativeness, of which the case $0<q<1$ and $1<q<2$ were analyzed in [36] and [37], respectively.

By substituting Eq. (2) into Eq. (A14) and introducing the variable substitution $t\rightarrow t+\tau$, the following integral is then yielded due to the linear property of the wedge product.

$$M_{F}\left(\tau\right)=\int_{-\infty}^{+\infty}\boldsymbol{f}\left(\boldsymbol{u}^{0}\left(t\right)\right)\wedge\boldsymbol{g}_{F}\left(\mathrm{D}^{q}\boldsymbol{u}^{0}\left(t\right),\boldsymbol{u}^{0}\left(t\right),t+\tau\right)\mathrm{d}t \tag{3}$$

In fact, $-\varepsilon M_{F}\left(\tau\right)$ is the measure of the splitting distance between $W^{\mathrm{s}}$ and $W^{\mathrm{u}}$ induced by the perturbation arising from the fractional-order element, and it is also the main thing to be noted in the Melnikov analysis of fractional-order systems.

The non-locality of fractional calculus defined by the convolutional form provides advantages for mathematical modeling of the history-dependent problems. However, the power-law weakly singular kernel also leads to difficulties in the Melnikov analysis within the existing works, to the extent that Melnikov criterion for fractional-order systems has long been considered 'cannot be obtained explicitly.' In existing works, it is generally stopped at the step of Eq. (3), turning instead to alternative approaches to address the problem. According to the information available to the authors, Melnikov method was first applied [22] to the global dynamics analysis of fractional-order systems in 2015. The relevant works [22-35] since then and to date are summarized in Table 1 here into two categories, according to the way to deal with the fractional-order perturbation described by Eq. (3).

As listed in Table 1, one of the approaches to deal with the problem is based on the equivalence principle [36]. This approach was first introduced in the study on a fractional-order

Duffing system [28], and then it was more adopted [29-35]. Generally, one can obtain an approximate analytical solution of periodic response based on perturbation methods, and thus find the equivalence principle of fractional-order elements. Then, the subsequent complex analysis can be ingeniously avoided by replacing the obtained equivalence principle. Due to the frequency-dependent characteristics of the equivalence principle, the only thing one should pay attention to is the restriction for frequency range introduced in the perturbation process of the approximate analytical solution, which ultimately cause a restriction for the applicable frequency range of the obtained Melnikov criterion. Nevertheless, within the frequency range of the approximate analytical solution, the Melnikov criterion obtained by this approach can not only be used to estimate the chaos threshold, but also qualitatively reflect the local evolution of the chaos threshold. Another approach [22-27] is to use numerical integration, and finally obtain a semi-analytical Melnikov criterion. This approach, in contrast to the former one, is not limited by the frequency range and therefore allows global numerical estimation, but it is generally difficult to draw qualitative conclusions.

Also in Table 1, the definitions of fractional-order derivative, the Hamiltonian of the unperturbed system, and the types of hyperbolic orbits studied in existing work are summarized by groups. The results also suggest some distinctions regarding the choice of definitions. Nevertheless, for a sufficiently smooth function $f(t)$, its Riemann-Liouville derivative and Grünwald-Letnikov derivative are equivalent [19]. Moreover, under certain restrictive initial conditions, the Riemann-Liouville derivative and the Caputo derivative are also equivalent [38]. These two are the basis for allowing different definitions to be employed for the Melnikov analysis of fractional-order systems listed in Table 1. In these works, however, Caputo definition is the most adopted because the derivative can be initialized using physically interpretable initial values and boundary conditions.

As a cross-disciplinary theme between qualitative theory and fractional dynamics, the global dynamics of various fractional-order systems with hyperbolic orbits have been extensively studied in the existing literature through semi-analytical or equivalence principle-based approaches. These studies have provided valuable and insightful perspectives on the chaotic dynamics of mechanical systems described by these models, such as the tri-stable energy harvesting system [22], tension leg platform system [23], nonlinear suspension system [25]. Nevertheless, there is a universal omission regarding the calculation of fractional-order perturbation that should be clarified.

### 2.2 Clarification and redefinition

As also indicated in Table 1, the left derivative is more frequently adopted in these studies. This is because, as can be learned from Appendix B, the left derivative for time actually describes the history-dependent property of the function, i.e., the fractional-order derivative of $f(t)$ at time $t$ depends on the history within $[a,t]$. Thus, the left derivative is more widely employed due to the fact that it is more suitable for modeling real-world physical phenomena. On the other hand, the right derivative describes the future-dependent property, i.e., the fractional-order derivative of $f(t)$ at time $t$ depends on the future within $[t,b]$. Such history- and future-dependent properties are also reflected in the necessity that the time boundaries should be specified when defining a fractional-order differential equation. However, whether the left or right derivative is employed, improperly defining their boundaries will lead to an incorrect estimation for $M_F(\tau)$ in Melnikov analysis.

In the case of left derivatives, for example, the left Caputo derivative, will generally lead to an integral from Eq. (3) as follows.

$$M_F(\tau) = \xi \cdot \int_{-\infty}^{+\infty} \int_{a}^{t} \Theta(s,t,t+\tau)\, \mathrm{d}s\ \mathrm{d}t \tag{4}$$

where $\Theta = \boldsymbol{f} \wedge \boldsymbol{g}_F$, which will be presented again in Section 3 with specific example, generally depend on the order of the derivative, the type and period of hyperbolic orbit. And $\xi$ describes those terms that depend on system parameters and are independent of time.

Furthermore, it can be learned from Eqs. (A10) to (A12) that the basic idea of Melnikov method is to use the globally computable solution of the unperturbed system in the perturbation calculations, for which it is necessary to ensure that the perturbation calculations validate at arbitrary time $t$. Therefore, in the calculation for fractional-order perturbation $M_F(\tau)$, **it is** **naturally required that the left derivative should be defined by** ${}^{L}\mathrm{D}_{-\infty,t}^{q}$ instead of ${}^{L}\mathrm{D}_{0,t}^{q}$ as in the existing works. Otherwise, it would result in that $\boldsymbol{d}_N^{\mathrm{u}}$, the splitting distance of the unstable manifold in Eq. (A12), is actually calculated in the following manner.

$$\boldsymbol{d}_N^{\mathrm{u}}(\tau,\tau)=\int_0^{\tau}\boldsymbol{f}\left(\boldsymbol{u}^0(t-\tau)\right)\wedge\boldsymbol{g}\left(\boldsymbol{u}^0(t-\tau)\right)\mathrm{d}t \tag{5}$$

This means that at least the perturbation calculations for unstable manifolds $W^{\mathrm{u}}$ lying on $(-\infty,0)$ are neglected.

In fact, defining boundary for the fractional-order element by any specific number $a$, i.e., ${}^{L}\mathrm{D}_{a,t}^{q}$, will result in the perturbation calculations for the entire $W^{\mathrm{u}}$ over $(-\infty,a)$, a part of $W^{\mathrm{u}}$ over $(a,\tau)$, and a part of $W^{\mathrm{s}}$ over $(\tau,+\infty)$ are neglected due to the fact that in Eq. (4), $\int_{-\infty}^{+\infty}\int_a^t\Theta(s,t,t+\tau)\,\mathrm{d}s\,\mathrm{d}t=\int_a^{+\infty}\int_a^t\Theta(s,t,t+\tau)\,\mathrm{d}s\,\mathrm{d}t$ for any function $\Theta(s,t)$ integrable over the region $\Omega=\{(s,t)\,|\,a<s<t\}\subset\mathbb{R}^2$. Similarly, **for the right derivative, the fractional-order element in Eq. (2) should be defined by** ${}^{R}\mathrm{D}_{t,+\infty}^{q}$. Otherwise, defining boundary for the fractional-order element by any specific number $b$, i.e., ${}^{R}\mathrm{D}_{t,b}^{q}$, will result in the perturbation calculations for the entire $W^{\mathrm{s}}$ over $(b,+\infty)$, a part of $W^{\mathrm{s}}$ over $(\tau,b)$, and a part of $W^{\mathrm{u}}$ over $(-\infty,\tau)$ are neglected.

As a visual illustration for the above suggestions, a graphical explanation is plotted in Fig. 1 using the integration region of perturbation $M_F(\tau)$ under the fractional-order element defined within a specific time boundary.

Note that both the Riemann-Liouville derivative and the Caputo derivative are based on the Riemann-Liouville integral defined by Eq. (A16), and their memory boundaries originate from this integral. Consequently, global perturbation calculation for Riemann-Liouville type fractional-order systems will, in fact, lead to Eq. (4) and subsequently to the following discussion, just like the Caputo derivative. Additionally, for the Grünwald-Letnikov type, if the system (1) is sufficiently smooth [19], which is precisely the case studied in this paper, Eq. (4)

will also yield due to the fact that ${}_{\mathrm{GL}}\mathrm{D}^{q} f\left(t\right) = {}_{\mathrm{RL}}\mathrm{D}^{q} f\left(t\right)$. In short, the discussion in this section applies to the three definitions of fractional-order derivatives introduced in Appendix B.

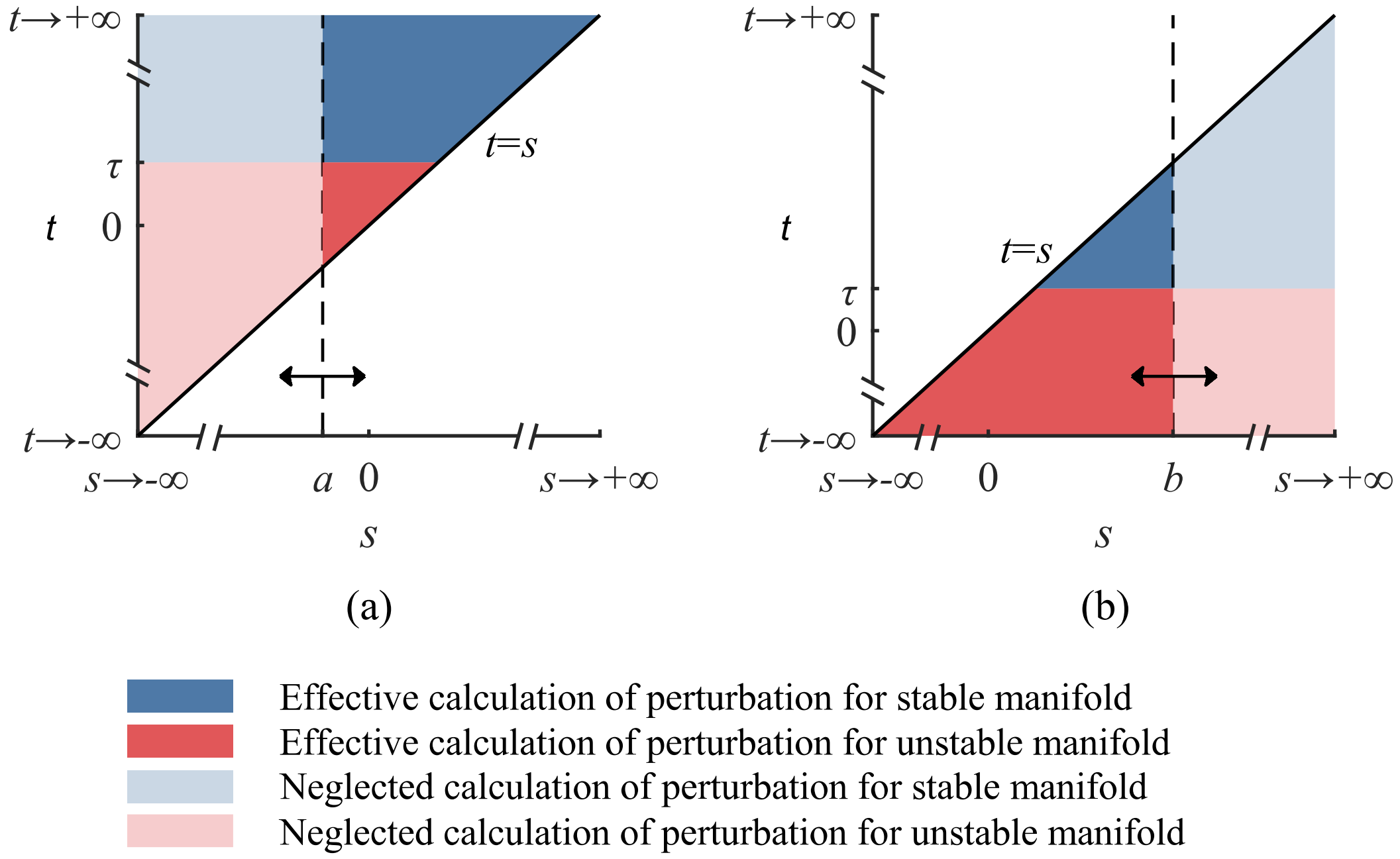

**Fig. 1.** Integration region of perturbation $M_F\left(\tau\right)$ under the fractional-order element defined within a specific time boundary. (a) The case of left derivative ${}^{L}\mathrm{D}_{a,t}^{q}$; (b) The case of right derivative ${}^{R}\mathrm{D}_{t,b}^{q}$.

**Table 1.** Summary of existing research on Melnikov analysis of fractional-order systems.

| Approach | Definition † | Hamiltonian ‡ | Hyperbolic orbit | Ref. |
|---|---|---|---|---|
| Semi-analytical | ${}^{R}_{\mathrm{C}}\mathrm{D}^{q}_{t,0}$ | $\frac{m}{2}x_2^2+\frac{k}{2}x_1^2-\frac{\alpha}{4}x_1^4+\frac{\beta}{6}x_1^6$ | Both homoclinic and heteroclinic | [22] |
| | ${}^{L}_{\mathrm{GL}}\mathrm{D}^{q}_{0,t}$ | $\frac{m}{2}x_2^2+\frac{k}{2}x_1^2+\frac{\alpha}{4}x_1^4-\frac{\beta}{6}x_1^6$ | Heteroclinic | [23] |
| | ${}^{L}_{\mathrm{C}}\mathrm{D}^{q}_{0,t}$ | $\frac{m}{2}x_2^2+\frac{k}{2}x_1^2-\frac{\alpha}{4}x_1^4$ | | [24, 25] |
| | | $\frac{m}{2}x_2^2-\frac{k}{2}x_1^2+\frac{\alpha}{4}x_1^4$ | Homoclinic | [26, 27] |
| Equivalence principle-based | | $\frac{m}{2}x_2^2-\frac{k}{2}x_1^2+\frac{\alpha}{4}x_1^4$ | | [28-31] |
| | | $\frac{m}{2}x_2^2-\frac{k}{2}x_1^2+\frac{\alpha}{\beta+2}x_1^2\lvert x_1\rvert^{\beta}$ | | [32] |
| | | $\frac{m}{2}x_2^2+\frac{k}{2}x_1^2-\frac{\alpha}{4}x_1^4$ | Heteroclinic | [33-35] |

† Note that although some of these works initially adopted definition ${}^{L}\mathrm{D}^{q}_{-\infty,t}$ or ${}^{R}\mathrm{D}^{q}_{t,+\infty}$, both the analysis and simulation therein were ultimately based on ${}^{L}\mathrm{D}^{q}_{a,t}$ or ${}^{R}\mathrm{D}^{q}_{t,b}$. Hence, the governing equations are in fact defined by the latter, which are listed below.

‡ Where $m$, $k$, $\alpha$, $\beta$ denote parameters of the studied system and are all positive.

## 3 A rigorous case of Melnikov analysis for fractional-order system

### 3.1 Dynamics of unperturbed system

At this point, it is clarified how to define the problem. Next, as a demonstration case, the following fractional-order Duffing-Rayleigh system [39] is used to re-introduce the Melnikov analysis process of fractional-order systems.

$$\ddot{x}+\bar{\mu}\left(1-\beta\dot{x}^2\right)\dot{x}+\bar{\gamma}\,{}_{\mathrm{C}}^{L}\mathrm{D}_{-\infty,t}^{q}x+\omega_0^2x-\alpha x^3=\bar{F}\cos\omega t \tag{6}$$

where $\alpha>0$ and $1<q<2$. Interest in the chaotic dynamics of this system arises from its application in communication, that is, the use of chaotic synchronization for encrypted signal transmission. The introduction of fractional-order elements is due to their non-locality, which is generally believed to effectively enhance the unpredictability of chaotic synchronization controllers.

By introducing the dimensionless bookkeeping parameter $\varepsilon$ and the parametric substitutions $\mu=\varepsilon\bar{\mu}$, $\gamma=\varepsilon\bar{\gamma}$ and $F=\varepsilon\bar{F}$, Eq. (6) can be rewritten in the form of Eq. (1) with

$$f(\boldsymbol{x})=\begin{bmatrix}x_2\\ \alpha x_1^3-\omega_0^2x_1\end{bmatrix} \tag{7a}$$

and

$$g(\boldsymbol{x},t)=\begin{bmatrix}0\\ F\cos\omega t-\mu\left(1-\beta x_2^2\right)x_2-\gamma\,{}_{\mathrm{C}}^{L}\mathrm{D}_{-\infty,t}^{q}x_1\end{bmatrix} \tag{7b}$$

Thus, the Hamiltonian of unperturbed system is given by

$$H\left(x_1,x_2\right)=\frac{1}{2}x_2^2+\frac{1}{2}\omega_0^2x_1^2-\frac{1}{4}\alpha x_1^4 \tag{8}$$

At its three equilibrium points $C:(0,0)$, $S_1:\left(-\frac{\omega_0}{\sqrt{\alpha}},0\right)$ and $S_2:\left(\frac{\omega_0}{\sqrt{\alpha}},0\right)$, the Hamiltonian of the unperturbed system is characterized as follows.

$$\frac{\partial H}{\partial x_1}=0,\ \frac{\partial H}{\partial x_2}=0,\ \frac{\partial^2 H}{\partial x_1^2}=-2\omega_0^2<0,\ \frac{\partial^2 H}{\partial x_2^2}=1>0\ \text{ at both } S_1 \text{ and } S_2 \tag{9a}$$

$$\frac{\partial H}{\partial x_1}=0,\ \frac{\partial H}{\partial x_2}=0,\ \frac{\partial^2 H}{\partial x_1^2}=\omega_0^2>0,\ \frac{\partial^2 H}{\partial x_2^2}=1>0\ \text{ at } C \tag{9b}$$

Therefore, both $S_1$ and $S_2$ are saddle equilibria and $C$ is the center equilibrium. As shown in Fig. 2, the heteroclinic orbit connecting $S_1$ and $S_2$ exists in the phase portrait of the unperturbed system, and it can be obtained by substituting the Hamiltonian at the saddle points into the differential relationship of the state variables $(x_1, x_2)^{\mathrm{T}}$ and then solving the differential equation as follows.

$$\boldsymbol{u}_{\pm}^{0}(t)=\begin{bmatrix} \pm\dfrac{\omega_0}{\sqrt{\alpha}}\tanh\left(\dfrac{\omega_0 t}{\sqrt{2}}\right) \\ \pm\dfrac{\omega_0^2}{\sqrt{2\alpha}}\operatorname{sech}^2\left(\dfrac{\omega_0 t}{\sqrt{2}}\right) \end{bmatrix} \tag{10}$$

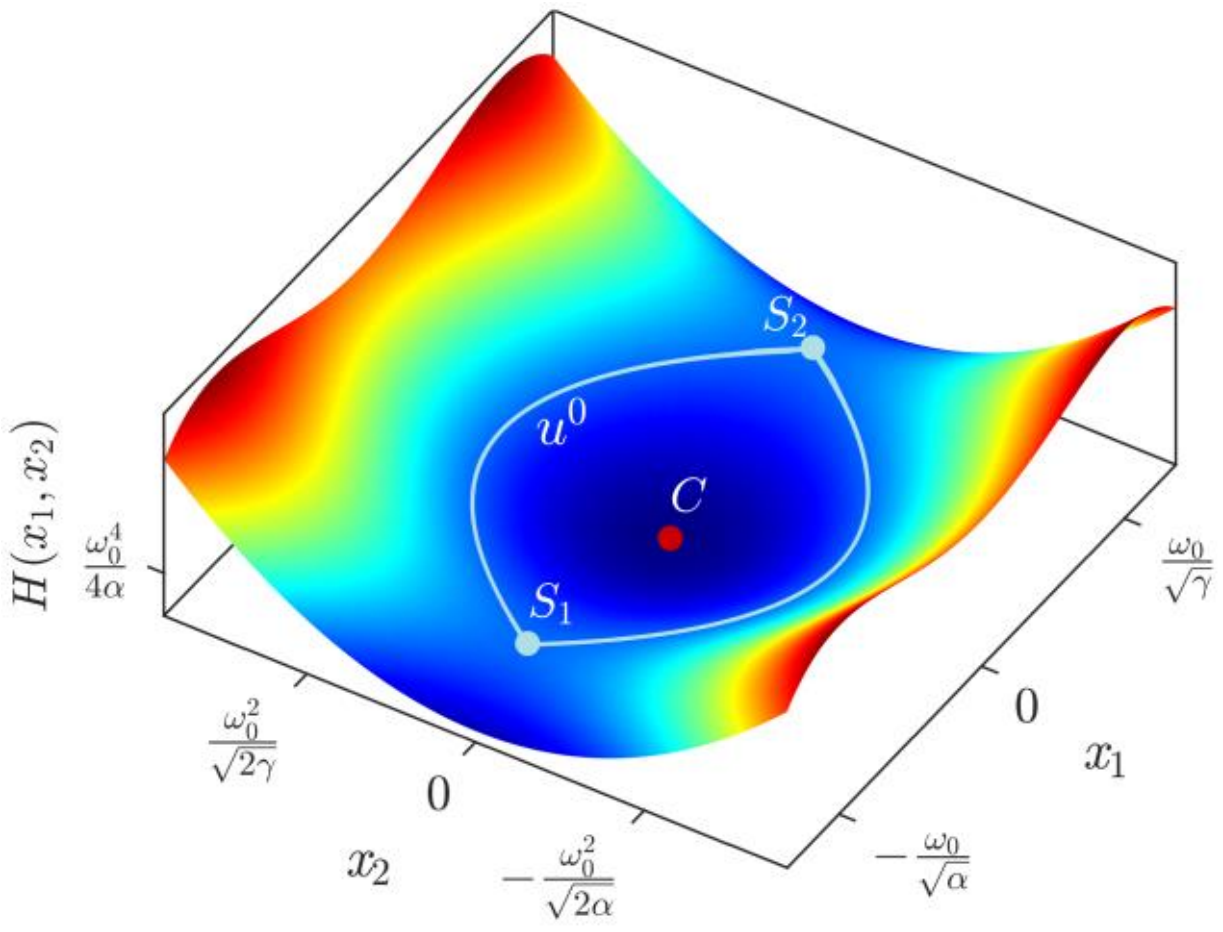

**Fig. 2.** Phase portrait of the unperturbed system.

### 3.2 Melnikov criterion of perturbated system

Based on the discussion in subsection 2.2 and Eq. (A14), Melnikov function of system (5) can be calculated as follows.

$$\begin{aligned} M(\tau) &= \int_{-\infty}^{+\infty} f\left[\boldsymbol{u}_{+}^{0}(t)\right]\wedge g\left[\boldsymbol{u}_{+}^{0}(t), t+\tau\right]\mathrm{d}t \\ &= \int_{-\infty}^{+\infty}\frac{\omega_0^2}{\sqrt{2\alpha}}\operatorname{sech}^2\left(\frac{\omega_0 t}{\sqrt{2}}\right)\cdot\left[F\cos\omega(t+\tau)-\frac{\mu\omega_0^2}{\sqrt{2\alpha}}\left(1-\frac{\beta\omega_0^4}{2\alpha}\operatorname{sech}^4\left(\frac{\omega_0 t}{\sqrt{2}}\right)\right)\operatorname{sech}^2\left(\frac{\omega_0 t}{\sqrt{2}}\right)-\frac{\gamma\omega_0}{\sqrt{\alpha}}\,{}_{\mathrm{C}}^{L}\mathrm{D}_{-\infty,t}^{q}\tanh\left(\frac{\omega_0 t}{\sqrt{2}}\right)\right]\mathrm{d}t \\ &= M_I(\tau)+M_F(\tau) \end{aligned} \tag{11}$$

where $\boldsymbol{u}_{+}^{0}(t)$ is used and similar for $\boldsymbol{u}_{-}^{0}(t)$.

$M_I(\tau)$, which represents all perturbations not involving fractional-order element, is integrated as follows.

$$\begin{aligned} M_I(\tau) &= \int_{-\infty}^{+\infty} \frac{\omega_0^2}{\sqrt{2\alpha}} \operatorname{sech}^2\left(\frac{\omega_0 t}{\sqrt{2}}\right) \cdot \left[ F\cos\omega(t+\tau) - \frac{\mu\omega_0^2}{\sqrt{2\alpha}}\left(1 - \frac{\beta\omega_0^4}{2\alpha}\operatorname{sech}^4\left(\frac{\omega_0 t}{\sqrt{2}}\right)\right)\operatorname{sech}^2\left(\frac{\omega_0 t}{\sqrt{2}}\right)\right]\mathrm{d}t \\ &= \frac{\sqrt{2}\pi\omega F}{\sqrt{\alpha}}\cos(\omega\tau)\operatorname{csch}\left(\frac{\pi\omega}{\sqrt{2}\omega_0}\right) - \frac{2\sqrt{2}\mu\omega_0^3}{3\alpha} + \frac{8\sqrt{2}\beta\mu\omega_0^7}{35\alpha^2} \end{aligned} \tag{12}$$

As for $M_F(\tau)$, the fractional-order perturbation, it will be treated here based on rigorous analysis instead of neither the semi-analytical nor equivalence principle-based approach summarized in Table 1. According to the discussion in subsection 2.2, it is suggested to be calculated in the following way.

$$\begin{aligned} M_F(\tau) &= \int_{-\infty}^{+\infty} \frac{\omega_0^2}{\sqrt{2\alpha}} \operatorname{sech}^2\left(\frac{\omega_0 t}{\sqrt{2}}\right) \cdot \left[ -\frac{\gamma\omega_0}{\sqrt{\alpha}} {}_{\mathrm{C}}^{L}\mathrm{D}_{-\infty,t}^{q} \tanh\left(\frac{\omega_0 t}{\sqrt{2}}\right)\right]\mathrm{d}t \\ &= \frac{\gamma\omega_0^5}{\sqrt{2\alpha}} \frac{1}{\Gamma(2-q)} I \end{aligned} \tag{13}$$

where $I$ is exactly the specific case of the integral in Eq. (4) under heteroclinic orbits and $1 < q < 2$, i.e.,

$$I = \int_{-\infty}^{+\infty} \operatorname{sech}^2\left(\frac{\omega_0 t}{\sqrt{2}}\right) \int_{-\infty}^{t} (t-s)^{1-q} \operatorname{sech}^2\left(\frac{\omega_0 s}{\sqrt{2}}\right) \tanh\left(\frac{\omega_0 s}{\sqrt{2}}\right) \mathrm{d}s\,\mathrm{d}t \tag{14}$$

First, rewrite the integral $I$ by introducing $p = 1-q$ and $\varphi = \frac{\omega_0}{\sqrt{2}}$, and then evaluate the inner integral:

$$J(t) = \int_{-\infty}^{t} (t-s)^p \tanh(\varphi s) \operatorname{sech}^2(\varphi s)\,\mathrm{d}s \tag{15}$$

Denote $\Psi$ as $\mathbb{C}$ takeout $t$ and all zeroes of $\cosh(\varphi s)$, and then define $\psi : \Psi \to \mathbb{C}$ via

$$\psi(s) = \mathrm{e}^{p\ln(t-s)} \tanh(\varphi s) \operatorname{sech}^2(\varphi s) \tag{16}$$

Note that $\psi$ is holomorphic away from $(-\infty, t]$. Then, by applying the residue theorem, $\psi(z)$ is integrated as follows.

$$\int_{\mathscr{H}} \psi(z) = 2\pi\mathrm{i} \sum_{\kappa\in\mathbb{Z}} \operatorname{Res}\left(\psi; \frac{\pi\mathrm{i}}{2\varphi}(2\kappa+1)\right) \tag{17}$$

where $\mathscr{H}$ stands for a Hankel contour that winds $-\infty+\mathrm{i}0^+ \to t+\mathrm{i}0^+ \to t+\mathrm{i}0^- \to -\infty+\mathrm{i}0^-$.

Thus, one can obtain $J(t)$ as follows.

$$J(t)=\pi\left(\cot(p\pi)-\mathrm{i}\right)\sum_{\kappa\in\mathbb{Z}}\operatorname{Res}\left(\psi;\frac{\pi\mathrm{i}}{2\varphi}(2\kappa+1)\right) \tag{18}$$

Then, by counting the residues at the poles, the inner integral $J(t)$ can be obtained as follows.

$$J(t)=-\pi\varphi^{-3}q(q-1)\csc(q\pi)\cdot\Re\left(\sum_{\kappa\in\mathbb{N}}\frac{\mathrm{e}^{p\ln\left(\frac{\pi\mathrm{i}}{2\varphi}(2\kappa+1)-t\right)}}{\left(t-\frac{\pi\mathrm{i}}{2\varphi}(2\kappa+1)\right)^2}\right) \tag{19}$$

where $\Re$ stands for the symmetric summation regarding i.

Then, $I$ continues to be evaluated by

$$\begin{aligned} I &= \int_{-\infty}^{+\infty}\operatorname{sech}^2(\varphi t)\cdot J(t)\,\mathrm{d}t \\ &= -\varphi^{q-3}\pi q(q-1)\cdot\csc(q\pi)\cdot\Re\left(\int_{-\infty}^{+\infty}\operatorname{sech}^2(t)\cdot\left(\sum_{\kappa\in\mathbb{N}}\mathrm{e}^{-(q+1)\ln\left(\frac{\pi\mathrm{i}}{2}(2\kappa+1)-t\right)}\right)\mathrm{d}t\right)\end{aligned} \tag{20}$$

Denote $\Lambda(t)=\operatorname{sech}^2(t)\cdot\left(\sum_{\kappa\in\mathbb{N}}\mathrm{e}^{-(q+1)\ln\left(\frac{\pi\mathrm{i}}{2}(2\kappa+1)-t\right)}\right)$ and $I_0=\int_{-\infty}^{+\infty}\Lambda(t)\,\mathrm{d}t$. Then by applying the residue theorem again yields

$$I_0=-2\pi\mathrm{i}\cdot\sum_{\lambda\in\mathbb{N}}\operatorname{Res}\left(\Lambda;-\frac{\pi\mathrm{i}}{2}(2\lambda+1)\right) \tag{21}$$

Finally, by counting the residues at the poles again, $I$ is obtained in the following closed form.

$$I=\frac{q\left(q^2-1\right)}{\pi^q\varphi^{3-q}}\cdot\sec\frac{q\pi}{2}\cdot\zeta(q+1) \tag{22}$$

where $\zeta(\bullet)$ stand for Riemann Zeta function.

Thus, by integrating Eqs. (12), (13), and (22), Melnikov function $M(\tau)$ of the fractional-order Duffing-Rayleigh system described by Eq. (6) is obtained in the following closed-form.

$$M(\tau) = \frac{\sqrt{2}\pi\omega F}{\sqrt{\alpha}}\cos(\omega\tau)\operatorname{csch}\left(\frac{\pi\omega}{\sqrt{2}\omega_0}\right) - \frac{2\sqrt{2}\mu\omega_0^3}{3\alpha} + \frac{8\sqrt{2}\beta\mu\omega_0^7}{35\alpha^2} + 2^{1-\frac{q}{2}}\frac{\omega_0^{2+q}\pi^{-q}\gamma q\left(q^2-1\right)\zeta(1+q)}{\alpha\Gamma(2-q)}\sec\left(\frac{q\pi}{2}\right) \tag{23}$$

Recalling the introduction in Appendix A, $M(\tau)$ is the measure of the distance between $W^{\mathrm{s}}$ and $W^{\mathrm{u}}$, and the transverse intersection of $W^{\mathrm{s}}$ and $W^{\mathrm{u}}$ is a sign of chaos, which requires $M(\tau)=0$ and $\mathrm{d}M(\tau)/\mathrm{d}\tau \neq 0$. The necessary condition for this to be reached is

$$F\pi\omega\operatorname{csch}\left(\frac{\pi\omega}{\sqrt{2}\omega_0}\right) > \frac{2\mu\omega_0^3}{3\sqrt{\alpha}} - \frac{8\beta\mu\omega_0^7}{35\alpha^{3/2}} - \frac{2^{1-\frac{q}{2}}\gamma\omega_0^{2+q}q\left(q^2-1\right)}{\sqrt{2\alpha}\,\pi^q}\cdot\frac{\zeta(1+q)}{\Gamma(2-q)}\cdot\sec\frac{q\pi}{2} \tag{24}$$

In general, Eq. (24) is referred to as the Melnikov criterion for Eq. (6), and the critical parameter values that satisfy this criterion are called the chaos threshold. This criterion, unlike existing ones, is obtained in a closed form, thus explicitly reflecting the evolution of the chaos threshold. As a validation, the following parameters, $\bar{\mu}=0.4$, $\beta=-0.01$, $\bar{\gamma}=0.1$, $\alpha=1$, $\omega=1.1$, will be used to calculate the chaos threshold and compare it with the result obtained by direct numerical integration for $M(\tau)$. Before this, as an example, the existence of chaotic response in the system with $q=1.1$ and $\omega_0=1$ is first confirmed. Under this set of parameters, the dynamic bifurcation of the system with varying excitation amplitude $\bar{F}$ is shown in Fig. 3(a). Additionally, Lyapunov exponent is generally considered a reliable quantitative indicator for identifying chaotic attractors, even in fractional-order systems [40-42]. Here, it is computed based on the method we proposed in a previous work [43], as shown in Fig. 3(b), and then used to supportively confirm the nature of the chaotic response. It is suggested by Fig. 3(a) that the system enters chaos via the route of period-doubling bifurcation. Also, the positive Lyapunov Exponent in Fig. 3(b) is a quantitative indicator of the local divergence characteristic of chaos. Thus, the existence of chaotic response in the system is confirmed. In addition, Fig. 3 shows that chaos appears as the excitation amplitude increases, just as qualitatively indicated by Eq. (24).

For further validation, Eq. (24) is used to calculate the analytical value of the chaos threshold for $\bar{F}$, which is then compared with the results by direct numerical integration for $M(\tau)$. Note that $q$ and $\omega_0$ are the only two system parameters involved in Eq. (14), and thus they dominate the subsequent analysis for fractional-order perturbations. In view of this, the comparison of the numerical and analytical results is performed in the $(q,\omega_0)$ parameter plane, which are shown in Fig. 4, and their relative errors are shown in Fig. 5. It is indicated by Figs. 4 and 5 that the analytical results are almost quantitatively consistent with the numerical results. Thus, the Melnikov criterion in Eq. (24) can provide an analytical prediction for the chaos threshold of the fractional-order Duffing-Rayleigh system described by Eq. (6). More importantly, as an example, this section has demonstrated how the Melnikov function of a fractional-order system can be obtained in a closed form after the problem was re-defined in subsection 2.2.

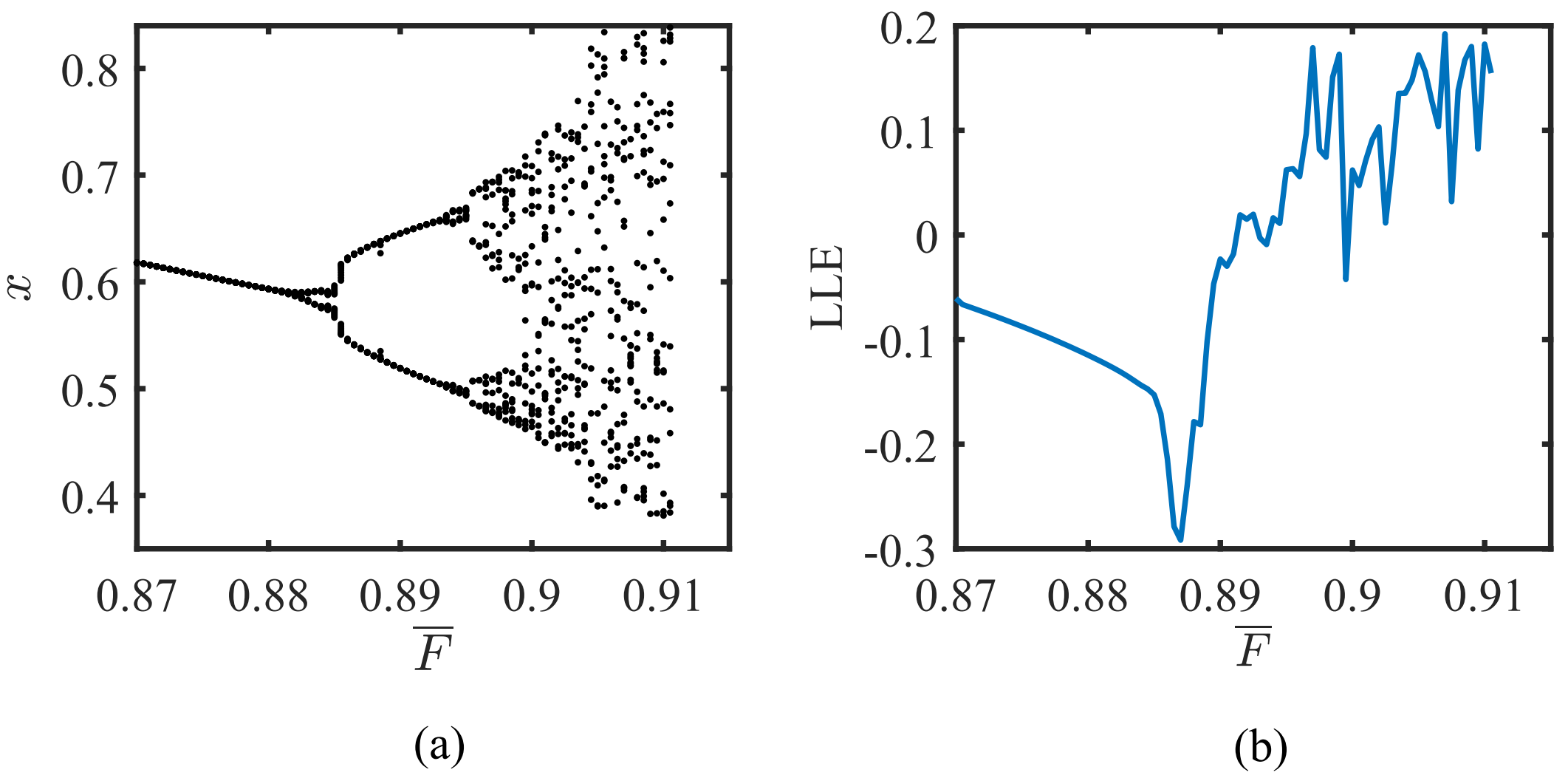

**Fig. 3.** Dynamics of the system for varying excitation amplitudes. (a) Period-doubling route to chaos; (b) The Largest non-trivial Lyapunov Exponent (LLE) computed by the memory principle-based method [43].

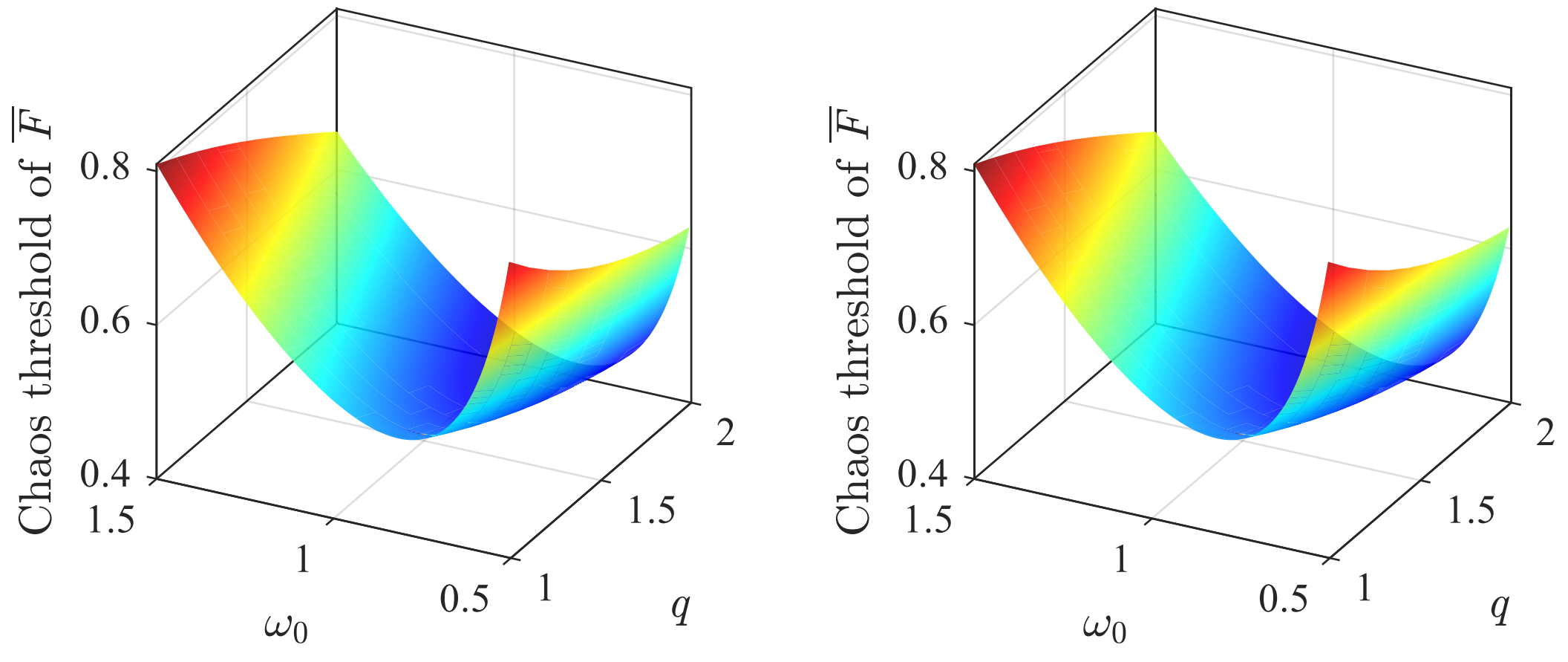

(a) (b)

**Fig. 4.** Verification of Melnikov criterion. (a) Chaos threshold calculated by Eq. (24); (b) Chaos threshold calculated by numerically integrating $M(\tau)$.

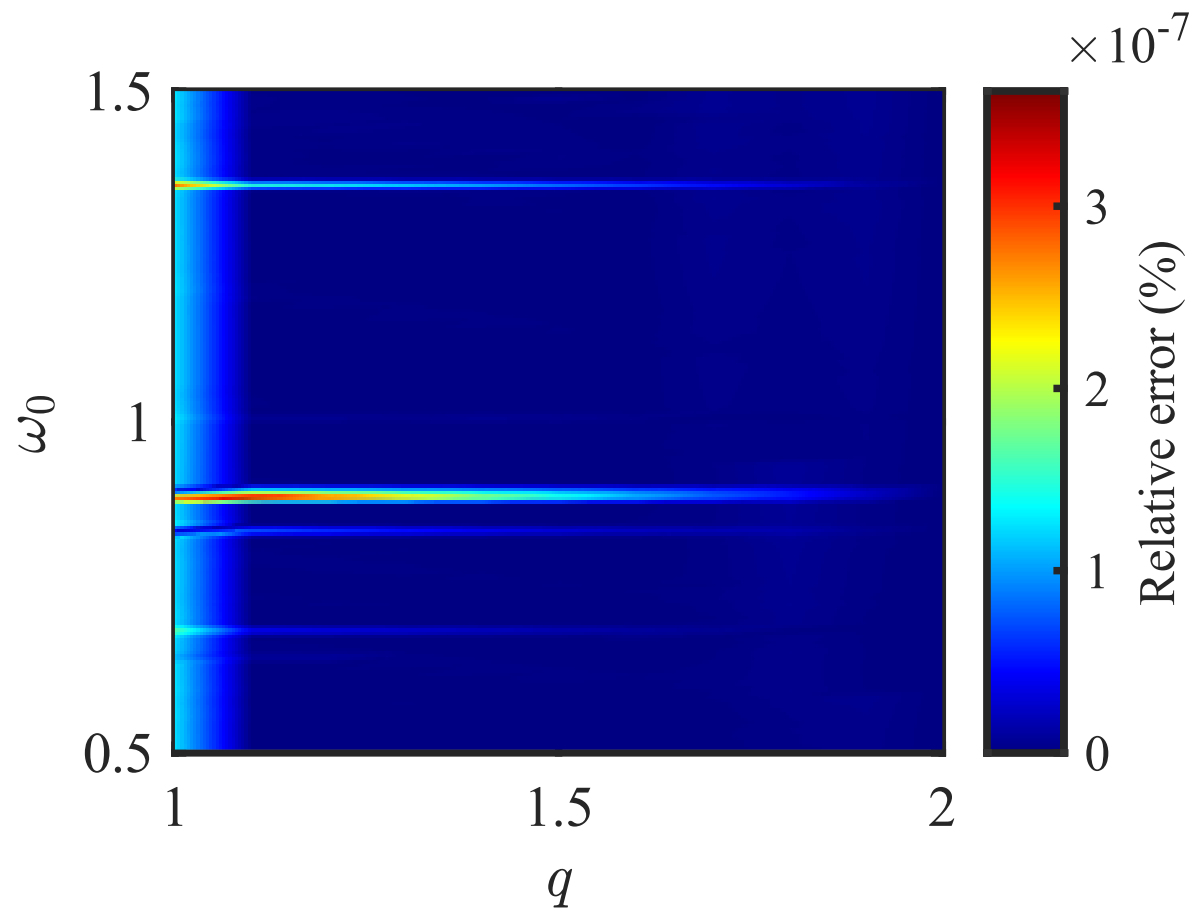

**Fig. 5.** Relative error between the analytical and numerical results for the chaos threshold.

## 4 Conclusions

In summary, Melnikov method for globally analyzing systems subjected to small time-varying periodic perturbation and the definition of fractional calculus are first introduced for subsequent discussion. In the governing equation, elements defined by fractional derivative should be considered as perturbation terms in the Melnikov analysis due to their non-conservative nature.

Then, the work based on this idea, i.e., Melnikov analysis of fractional-order systems, over the past decade is reviewed, summarized and categorized. Importantly, a universal omission therein is clarified. Specifically, whether the left derivative or the right derivative is adopted, the time boundaries of fractional-order elements need to be predefined due to the memory effect of time fractional-order derivatives. To ensure that the global perturbation analysis is uniformly valid, the fractional-order elements should be defined within infinite boundaries, i.e., ${}^{L}\mathrm{D}^{q}_{-\infty,t}$ or ${}^{R}\mathrm{D}^{q}_{t,+\infty}$. Otherwise, any definition within finite boundaries will result in the neglect of perturbation calculations for parts of the stable and unstable manifolds.

After clarifying and redefining the problem, the fractional-order Duffing-Rayleigh system is employed as a specific example to re-demonstrate the application of the Melnikov method in fractional dynamics. The Melnikov criterion is finally obtained in a closed form, different from ones obtained by the semi-analytical and equivalence principle-based approaches. At last, under the given demonstration parameters, the analytical value of the chaos threshold quantitatively matches the numerical value obtained by numerically integrating Melnikov function $M(\tau)$. Thus, the analysis is ultimately validated. The presented clarification and analysis are expected to provide insights into the topic of global dynamics of fractional-order systems.

**Declaration of interests**

The authors declare that they have no known competing financial interests or personal relationships that could have appeared to influence the work reported in this paper.

**Data Availability**

All data that support the findings of this work are available from the corresponding author upon reasonable request.

**CRediT authorship contribution statement**

**Hang Li**: Conceptualization, Methodology, Formal analysis, Investigation, Writing – original draft. **Yongjun Shen**: Methodology, Funding acquisition, Supervision, Project administration, Writing – review & editing. **Jian Li**: Conceptualization, Funding acquisition, Supervision, Project administration, Writing – review & editing. **Jinlu Dong**: Software, Validation, Formal analysis. **Guangyang Hong**: Funding acquisition, Visualization, Validation.

## Acknowledgments

This work is supported by the National Natural Science Foundation of China (Grant Nos. 12272091, 12272242 and 12302510), China Postdoctoral Science Foundation (No. 2023M740549), and the Fundamental Research Funds for the Central Universities (No. N2305015). Hang Li would like to acknowledge, in particular, Mr. Nathan from University of Oxford for his great help and insightful discussions in complex analysis.

## Appendix A

To facilitate discussion of details in the text, the Melnikov method is thoroughly introduced below. Assuming that in Eq. (1), $\boldsymbol{g}$ is characterized by a fixed time period $T$ and $0<|\varepsilon|\ll 1$. Thus, $\varepsilon\boldsymbol{g}$ can be regarded as a $O(\varepsilon)$ periodic perturbation for the following Hamiltonian system.

$$\dot{\boldsymbol{x}} = \boldsymbol{f}(\boldsymbol{x}) \tag{A1}$$

Then, assuming that a homoclinic orbit $\boldsymbol{u}^0(t)$ with hyperbolic saddle point $\boldsymbol{p}_0$ exist in the unperturbed system (A1). Then, define a Poincaré map $\boldsymbol{\Phi}_\varepsilon^\tau : \Sigma^\tau \to \Sigma^\tau$ at a fixed $\tau \in [0,T]$, where the Poincaré section $\Sigma^\tau \overset{\text{def}}{=} \left\{(\boldsymbol{x},\theta) \,\middle|\, \theta = \frac{2\pi\tau}{T} \in [0,2\pi]\right\}$. For the unperturbed system (A1), the saddle equilibrium point $\boldsymbol{p}_0$ is also a fixed point on the map $\boldsymbol{\Phi}_0^\tau$; for the perturbed system (1), according to the implicit function theorem, $\boldsymbol{\Phi}_\varepsilon^\tau$ also has a hyperbolic saddle point near $\boldsymbol{p}_0$, that is, $\boldsymbol{p}_\varepsilon^\tau = \boldsymbol{p}_0 + O(\varepsilon)$, and the corresponding homoclinic orbit near $\boldsymbol{p}_\varepsilon^\tau$ evolves into $\gamma_\varepsilon^0(t) = \boldsymbol{u}^0(t) + O(\varepsilon)$.

The stable manifolds $W^{\mathrm{s}}$ and unstable manifolds $W^{\mathrm{u}}$, where the orbit $\boldsymbol{u}^0$ lies, will split due to the perturbation. The orbit that originates from $\Sigma^\tau$ and lies on the split invariant manifolds can be uniformly expressed near $\boldsymbol{u}^0$ as follows.

$$\boldsymbol{r}_\varepsilon^{\mathrm{u}}(t,\tau)=\boldsymbol{u}^0(t-\tau)+\varepsilon\boldsymbol{u}_1^{\mathrm{u}}(t,\tau)+O(\varepsilon^2),\quad t\in(-\infty,\tau] \tag{A2a}$$

$$\boldsymbol{r}_\varepsilon^{\mathrm{s}}(t,\tau)=\boldsymbol{u}^0(t-\tau)+\varepsilon\boldsymbol{u}_1^{\mathrm{s}}(t,\tau)+O(\varepsilon^2),\quad t\in[\tau,+\infty) \tag{A2b}$$

Thus, the distance between $W^{\mathrm{s}}$ and $W^{\mathrm{u}}$ at time $t$ can be measured by

$$\begin{aligned}\boldsymbol{d}(t,\tau)&=\boldsymbol{r}_\varepsilon^{\mathrm{s}}(t,\tau)-\boldsymbol{r}_\varepsilon^{\mathrm{u}}(t,\tau)\\&=\varepsilon\left(\boldsymbol{u}_1^{\mathrm{s}}(t,\tau)-\boldsymbol{u}_1^{\mathrm{u}}(t,\tau)\right)+O(\varepsilon^2)\end{aligned} \tag{A3}$$

Denote $\boldsymbol{N}(t,\tau)$ as the normal vector of the orbit $\boldsymbol{u}^0$ at the point $\boldsymbol{u}^0(t-\tau)$. It can be determined as follows.

$$\boldsymbol{N}(t,\tau)=\left[-f_2\left(\boldsymbol{u}^0(t-\tau)\right),\, f_1\left(\boldsymbol{u}^0(t-\tau)\right)\right]^{\mathrm{T}} \tag{A4}$$

Then, by projecting $\boldsymbol{d}(t,\tau)$ onto $\boldsymbol{N}(t,\tau)$, the normal distance between $W^{\mathrm{s}}$ and $W^{\mathrm{u}}$ can be calculated by

$$\boldsymbol{d}_N(t,\tau)=\boldsymbol{N}\cdot\boldsymbol{d}=\boldsymbol{f}\wedge\boldsymbol{d}=\varepsilon\left(\boldsymbol{d}_N^{\mathrm{s}}-\boldsymbol{d}_N^{\mathrm{u}}\right)+O(\varepsilon^2) \tag{A5}$$

where $\boldsymbol{d}_N^{\mathrm{u}}$ and $\boldsymbol{d}_N^{\mathrm{s}}$, respectively, measure the splitting distance of the unstable manifold lying on $(-\infty,\tau]$ and the stable manifold lying on $(\tau,+\infty)$ relative to the unperturbed manifolds, and they are defined as follows.

$$\boldsymbol{d}_N^{\mathrm{u}}(t,\tau)\overset{\mathrm{def}}{=}\boldsymbol{f}\left(\boldsymbol{u}^0(t-\tau)\right)\wedge\boldsymbol{u}_1^{\mathrm{u}}(t,\tau) \tag{A6a}$$

$$\boldsymbol{d}_N^{\mathrm{s}}(t,\tau)\overset{\mathrm{def}}{=}\boldsymbol{f}\left(\boldsymbol{u}^0(t-\tau)\right)\wedge\boldsymbol{u}_1^{\mathrm{s}}(t,\tau) \tag{A6b}$$

Deriving both sides of Eq. (A6a) yields the following differential equation.

$$\dot{\boldsymbol{d}}_N^{\mathrm{u}}=\dot{\boldsymbol{f}}\wedge\boldsymbol{u}_1^{\mathrm{u}}+\boldsymbol{f}\wedge\dot{\boldsymbol{u}}_1^{\mathrm{u}}=\mathrm{D}\boldsymbol{f}\cdot\boldsymbol{u}^0\wedge\boldsymbol{u}_1^{\mathrm{u}}+\boldsymbol{f}\wedge\dot{\boldsymbol{u}}_1^{\mathrm{u}} \tag{A7}$$

where $\dot{\boldsymbol{u}}_1^{\mathrm{u}}$ can be calculated as follows by substituting Eq. (A2a) into Eq. (1).

$$\dot{\boldsymbol{u}}_1^{\mathrm{u}}(t,\tau)=\mathrm{D}\boldsymbol{f}\left(\boldsymbol{u}^0(t-\tau)\right)\boldsymbol{u}_1^{\mathrm{u}}(t,\tau)+g\left(\boldsymbol{u}^0(t-\tau),t\right),\quad t\in(-\infty,\tau] \tag{A8}$$

Hence Eq. (A7) is organized in the following form.

$$\dot{\boldsymbol{d}}_N^{\mathrm{u}}(t,\tau)=\mathrm{tr}\left[\mathrm{D}\boldsymbol{f}\left(\boldsymbol{u}^0(t-\tau)\right)\right]\boldsymbol{d}_N^{\mathrm{u}}(t,\tau)+\boldsymbol{f}\left(\boldsymbol{u}^0(t-\tau)\right)\wedge\boldsymbol{g}\left(\boldsymbol{u}^0(t-\tau)\right) \tag{A9}$$

Note that in Eq. (A9), $\mathrm{tr}\left[\mathrm{D}\boldsymbol{f}\left(\boldsymbol{u}^0\right)\right]=\mathrm{div}\left[\boldsymbol{f}\left(\boldsymbol{u}^0\right)\right]=0$ as the unperturbed system (A1) is conservative. Then, integrating Eq. (A9) over $t \in (-\infty, \tau]$ yields

$$\boldsymbol{d}_N^{\mathrm{u}}(\tau,\tau) - \boldsymbol{d}_N^{\mathrm{u}}(-\infty,\tau) = \int_{-\infty}^{\tau} \boldsymbol{f}\left(\boldsymbol{u}^0(t-\tau)\right) \wedge \boldsymbol{g}\left(\boldsymbol{u}^0(t-\tau)\right) \mathrm{d}t \tag{A10}$$

Noting that at $t = -\infty$, the orbit $\boldsymbol{r}_\varepsilon^{\mathrm{u}}(t,\tau)$ lying on $W^{\mathrm{u}}$ is located at the saddle equilibrium point $\boldsymbol{p}_\varepsilon^\tau$. Therefore, one has, according to Eq. (A6b), that

$$\boldsymbol{d}_N^{\mathrm{u}}(-\infty,\tau) = \boldsymbol{f}\left(\boldsymbol{p}_\varepsilon^\tau\right) \wedge \boldsymbol{u}_1^{\mathrm{u}}(t,\tau) = 0 \tag{A11}$$

Thus, $\boldsymbol{d}_N^{\mathrm{u}}$ is derived as follows.

$$\boldsymbol{d}_N^{\mathrm{u}}(\tau,\tau) = \int_{-\infty}^{\tau} \boldsymbol{f}\left(\boldsymbol{u}^0(t-\tau)\right) \wedge \boldsymbol{g}\left(\boldsymbol{u}^0(t-\tau)\right) \mathrm{d}t \tag{A12a}$$

Similarly for Eq. (A6b), $\boldsymbol{d}_N^{\mathrm{s}}$ can be derived as follows.

$$\boldsymbol{d}_N^{\mathrm{s}}(\tau,\tau) = -\int_{\tau}^{+\infty} \boldsymbol{f}\left(\boldsymbol{u}^0(t-\tau)\right) \wedge \boldsymbol{g}\left(\boldsymbol{u}^0(t-\tau)\right) \mathrm{d}t \tag{A12b}$$

Ultimately, by substituting Eq. (A12) into Eq. (A5) and neglecting higher-order terms, the normal distance between $W^{\mathrm{s}}$ and $W^{\mathrm{u}}$ can be approximated as follows.

$$\boldsymbol{d}_N(\tau,\tau) = -\varepsilon M(\tau) \tag{A13}$$

where $M(\tau)$ is defined as Melnikov function of perturbed system (1), which is formulated as follows.

$$M(\tau) = \int_{-\infty}^{+\infty} \boldsymbol{f}\left(\boldsymbol{u}^0(t-\tau)\right) \wedge \boldsymbol{g}\left(\boldsymbol{u}^0(t-\tau), t\right) \mathrm{d}t \tag{A14}$$

To summarize, $M(\tau)$ measures the splitting distance between $W^{\mathrm{s}}$ and $W^{\mathrm{u}}$, and it can be used to predict the formation of transverse homoclinic/heteroclinic points, which is the signature of the existence of the horseshoe map and its associated chaotic dynamics [44]. This analytical approach of predicting chaos by detecting the presence of transverse homoclinic/heteroclinic points is known as the Melnikov method.

**Appendix B**

For a given function $f(t)$ defined on $t \in (a,b)$, its $q$-order ($q > 0$) fractional derivative and integral can be defined [17, 19] in the following several ways.

**Definition 1** The $q$-order Grünwald-Letnikov derivative of $f(t)$ is defined by

$$ {}_{\mathrm{GL}}^{L}\mathrm{D}_{a,t}^{q} f(t) = \lim_{h\to 0} \frac{1}{h^q} \sum_{j=0}^{\left[\frac{t-a}{h}\right]} (-1)^j \binom{q}{j} f(t-jh) \tag{A15a} $$

or

$$ {}_{\mathrm{GL}}^{R}\mathrm{D}_{t,b}^{q} f(t) = \lim_{h\to 0} \frac{1}{h^q} \sum_{j=0}^{\left[\frac{b-t}{h}\right]} (-1)^j \binom{q}{j} f(t+jh) \tag{A15b} $$

where $(-1)^j \binom{q}{j}$ is the fractional binomial coefficient, and the operator $[\bullet]$ denotes taking the integer part. The upper left subscripts $L$ and $R$ of the derivative operator $D$ represent the left derivative and the right derivative respectively, and the same below.

**Definition 2** The $q$-order Riemann-Liouville integral of $f(t)$ is defined by

$$ {}_{\mathrm{RL}}^{L}\mathrm{I}_{a,t}^{q} f(t) = \frac{1}{\Gamma(q)} \int_a^t \frac{f(s)}{(t-s)^{1-q}} \mathrm{d}s \tag{A16a} $$

or

$$ {}_{\mathrm{RL}}^{R}\mathrm{I}_{t,b}^{q} f(t) = \frac{1}{\Gamma(q)} \int_t^b \frac{f(s)}{(s-t)^{1-q}} \mathrm{d}s \tag{A16b} $$

where $\Gamma(\bullet)$ is Euler Gamma function.

**Definition 3** The $q$-order Riemann-Liouville derivative of $f(t)$ is defined by

$$ {}_{\mathrm{RL}}^{L}\mathrm{D}_{a,t}^{q} f(t) = \frac{1}{\Gamma(n-q)} \frac{\mathrm{d}^n}{\mathrm{d}t^n} \int_a^t \frac{f(s)}{(t-s)^{q-n+1}} \mathrm{d}s \tag{A17a} $$

or

$$ {}_{\mathrm{RL}}^{R}\mathrm{D}_{t,b}^{q} f(t) = \frac{1}{\Gamma(n-q)} \frac{\mathrm{d}^n}{\mathrm{d}t^n} \int_t^b \frac{f(s)}{(s-t)^{q-n+1}} \mathrm{d}s \tag{A17b} $$

where $n = \lceil q \rceil$, and the same below.

**Definition 4** The $q$-order Caputo derivative of $f(t)$ is defined by

$$ {}_{\mathrm{C}}^{L}\mathrm{D}_{a,t}^{q} f(t) = \frac{1}{\Gamma(n-q)} \int_a^t \frac{f^{(n)}(s)}{(t-s)^{q-n+1}} \mathrm{d}s \tag{A18a} $$

or

$$
{}_{\mathrm{C}}^{R}\mathrm{D}_{t,b}^{q} f(t) = \frac{1}{\Gamma(n-q)} \int_t^b \frac{f^{(n)}(s)}{(s-t)^{q-n+1}} \mathrm{d}s \tag{A18b}
$$

In fact, there are many other definitions of fractional calculus. Considering the topic of this paper and the use of these definitions in the works summarized in Table 1, only the above definitions are presented here. For other definitions of fractional calculus and more details, one can refer to the works by Podlubny [17] and Li et al [19].